\documentclass[
  ,draft            
  ]
  {aipproc}

\layoutstyle{8x11double}

\begin{document}

\title{Critical Metallicities for Second-Generation Stars}


\author{J. Michael Shull}{
  address={Department of Astrophysical and Planetary Sciences, 
        University of Colorado, Boulder CO 80309, USA \\
          e-mail: mshull@casa.colorado.edu} 
}

\begin{abstract}
The first massive stars may influence the formation of second-generation 
stars, in part by their metal enrichment of the surrounding gas. We 
investigate the ``critical metallicity", defined as the the value, 
$Z_{\rm crit}$, at which primordial gas cools more efficiently by 
fine-structure lines of O~I (63.18 $\mu$m), Si~II (34.8 $\mu$m), 
Fe~II (25.99 and 35.35 $\mu$m), and C~II (157.74 $\mu$m) than by either 
H$^{\circ}$ or H$_2$ line emission.  We explore the time-dependent 
thermodynamics and fragmentation of cooling gas at redshifts 
$z = 10-30$, seeded by trace heavy elements expelled from early 
supernovae.  Because different modes of nucleosynthesis 
($\alpha$-process, Fe-group) produce abundance ratios far from 
solar values, these early stellar populations are likely to be 
influenced by O, Si, and Fe cooling.  Our models also include 
radiative coupling of the fine-structure lines and H$_2$ to the 
cosmic microwave background (CMB), which sets a temperature floor 
(70--80~K at $z =$ 25--30) that may increase the Jeans mass. The 
H$_2$ forms from catalytic effects of electrons left over from the 
recombination epoch or produced 
during virialization. These electrons form the H$^-$ ion 
(H$^{\circ}$ + e$^- \rightarrow$ H$^-$ + $\gamma$), which in turn forms 
H$_2$ through associative detachment (H$^-$ + H$^{\circ} 
\rightarrow$ H$_2$ + e$^-$). In virialized halos at $z = 10-30$, the 
gas densities ($n \approx 1-100$ cm$^{-3}$) are well below the critical 
densities, $n_{\rm cr} = 10^{5-6}$ cm$^{-3}$, at which (O, Si, Fe) 
fine-structure lines reach LTE populations and produce their most 
efficient cooling.  Thus, $Z_{\rm crit}$ may initially exceed 
$0.01 Z_{\odot}$ at $n \approx$ 1--100 cm$^{-3}$, and then drop to 
$10^{-3.5} Z_{\odot}$ at $n \sim 10^6$ cm$^{-3}$, where the Jeans 
mass may be imprinted on the stellar mass function.
Primordial clouds of $10^8~M_{\odot}$ at $0.01Z_{\odot}$ and 200~K 
will produce redshifted fine-structure lines, with fluxes between 
$10^{-22}$ and $10^{-21}$ W~m$^{-2}$ at $z \approx 4$.    
\end{abstract}

\maketitle


\section{Introduction}

What is meant by the term {\it Critical Metallicity} in the context 
of first- and second-generation stars?  We define 
$Z_{\rm crit}$ as the heavy-element abundance at which metal-line 
cooling of the gas begins to dominate over cooling by H and He
(and molecules H$_2$ and HD).  In order for the gas to collapse gravitationally 
and continue to radiate away the heat produced by adiabatic compression, 
the radiative cooling time,
$t_{\rm cool} \approx (3nkT/2 {\cal L})$, must be less than the 
gravitational collapse time, $t_{\rm coll} \approx 
(3 \pi/32 G \rho)^{1/2}$, where ${\cal L}$ is the cooling rate per 
volume and $\rho$ is the gas mass density. Recent studies,
based on Jeans-mass arguments and thermodynamic histories of cloud
collapse, suggest that the mode of star formation may change, 
shifting from higher-mass stars at low-$Z$ (Pop~III) to a 
more normal (Pop~II) initial mass function (IMF) at $Z > Z_{\rm crit}$. 

At zero metallicity, because of
the lack of CNO-burning and the inefficiency of the p-p chain, the
first stars are smaller, hotter, and shorter-lived than their current
counterparts (e.g., Tumlinson \& Shull 2000).  Their efficient production 
rates of ionizing radiation give them special importance for IGM 
reionization (Venkatesan, Tumlinson, \& Shull 2003; Wyithe \& Loeb
2003; Shull \& Venkatesan 2007). The most massive stars have large yields 
of heavy elements (Heger \& Woosley 2002), particularly $\alpha$-process
elements (O, Si) and the iron-group.  Thus, it is astrophysically
important to understand the transition from first to second-generation
stars when $Z > Z_{\rm crit}$.  However, this transition probably varies 
spatially and temporally, owing to the inhomogenous nature of metal 
production and transport into the IGM. 

In cold dark matter (CDM) cosmologies, the first galaxies in the
universe are predicted (Ricotti, Gnedin, \& Shull 2002, 2007)
to be $10^6$ times smaller than
the Milky Way, with characteristic masses comparable to mass estimates
for the smaller dwarf spheroidal galaxies (dSph) observed around our
Galaxy and Andromeda (Mateo 1998; Belokurov et~al.\ 2006). The
gravitational potentials of these $10^{6-8}~M_{\odot}$ objects are so weak
that warm and hot ionized phases of their interstellar medium
are weakly bound. As a result, each episode of star formation
may produce powerful outflows that could temporarily inhibit further
star formation.

The first subgalactic structures form from the collapse
of rare dark matter density perturbations, with masses
$10^{5-6}\,M_{\odot}$ at $z \sim 30-40$.  The initial gas cooling is 
from collisionally excited H$_2$ rotational and vibrational
transitions (Lepp \& Shull 1984).  A minimum H$_2$ abundance $x_{H_2}
\approx 10^{-4}$ is required to trigger star formation in a dark halo
in less than a Hubble time.  In dust-free gas, H$_2$ formation is
catalyzed by the H$^-$ ion, that forms as a consequence of the shocks
that partially ionize and heat the gas during the virialization
process. At a given redshift, the mass of the smaller halo that can
form stars is determined by its virial temperature and therefore by
its mass. This analytical result has been confirmed by hydrodynamical
cosmological simulations.

Abel, Bryan, \& Norman (2002) carried out
such numerical simulations for a selected $10^6$ M$_\odot$ halo, using
adaptive-mesh refinement that resolves the collapse over a large range of
scales.  In this selected halo, they find that only one star forms,
with mass 10--100 $M_{\odot}$.  Bromm, Coppi, \& Larsen
(1999) found similar results with a variety of initial conditions for
the protogalaxies.  These numerical results confirm longstanding
theoretical suggestions that the first stars should be
massive: their characteristic mass reflects the larger Jeans mass in
the inefficiently cooling metal-free gas.  However, the cooling by
trace-metal fine-structure lines depends on the gas density
(Santoro \& Shull 2006, 2008) and radiative coupling to the CMB.
Thus, the Jeans mass and critical metallicity are sensitive to 
local gas density.

\section{Previous Calculations of $Z_{\rm crit}$ }  

The first static models of $Z_{\rm crit}$ (Bromm \& Loeb 2003) found 
$Z_{\rm crit} \approx 10^{-3.5} Z_{\odot}$ for fine-structure cooling
by [C~II] 158~$\mu$m and [O~I] 63~$\mu$m.
Santoro \& Shull (2006) confirmed these results for C~II and O~I, 
but suggested that fine-structure lines of [Si~II] and [Fe~II]   
might also contribute, since IGM ``metal pollution" from massive-star 
nucleosynthesis is weighted toward heavier elements.  They further 
noted that $Z_{\rm crit}$
depends on the gas density, $n$, owing to the change in cooling 
rate, from ${\cal L} \propto n^2$ (at low-density) to 
${\cal L} \propto n$ (high-density), as the fine-structure levels 
reach Boltzmann (LTE) populations at ``critical density" 
($n_{\rm cr}$). For H$^{\circ}$ excitation at 200~K, $n_{\rm cr}$ ranges 
from $3000$ cm$^{-3}$ ([C~II]) to $(1-2) \times 10^6$ cm$^{-3}$ ([O~I]
and [Fe~II]). Santoro \& Shull (2006) found that $Z_{\rm crit}$ 
exceeds $0.01Z_{\odot}$ at the low gas densities,
$n \approx$ 1--100~cm$^{-3}$, present in virialized halos at $z > 20$.


\begin{figure}
  \includegraphics[height=10cm]{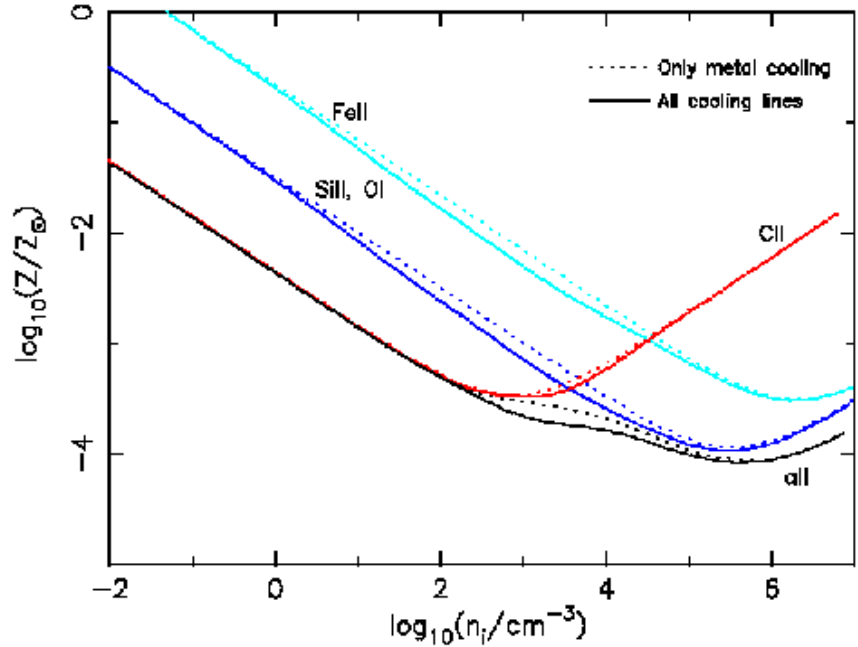}
  \caption{Minimum critical metallicities, $Z_{\rm crit}$,
   vs.\ total gas density $n$ (Santoro \& Shull 2006) 
   for static cooling at $T = 200$~K  by individual heavy elements.  
   Curves correspond to gas enriched by C~II, Si~II, O~I, and Fe~II
   (see labels).  Bottom envelope shows all four species together 
   in solar abundance ratios.  Minimum values occur at high densities
   (near $n_{\rm cr}$ for each coolant) 
   at log~$(Z_{\rm crit}/Z_{\odot}) = -3.48$ (C~II), $-3.54$ (Si~II), 
   $-3.78$ (O~I), $-3.52$ (Fe~II), and $-4.08$ (all elements). }
\end{figure}

Figure 1 shows values of $Z_{\rm crit}$ for C, O, Si, and Fe,
where metallicity is labelled by a single parameter $Z$.  In fact,
there is no single ``metallicity", since the primary coolants
(C, O, Si, Fe) are rarely produced in solar abundance ratios.
Because of the density dependence of the cooling, Santoro \& Shull 
(2008) examined the thermodynamic history of cloud collapse 
and refined the definition of $Z_{\rm crit}$.  In the new models
they included collisional coupling of the gas and level populations 
(H$_2$, HD, fine-structure lines) and radiative coupling 
to the cosmic microwave background (CMB).  The latter effect can be 
especially important at high redshifts, $z = 25-30$, where the CMB 
temperature, $T_{\rm CMB} = (82~K)[(1+z)/30]$, sets a floor on gas 
temperature sufficient to increase the Jeans mass.  This, in turn, 
may produce more massive stars at high redshift (Tumlinson 2007).   

\section{Results}


\begin{figure}
\includegraphics[height=12cm]{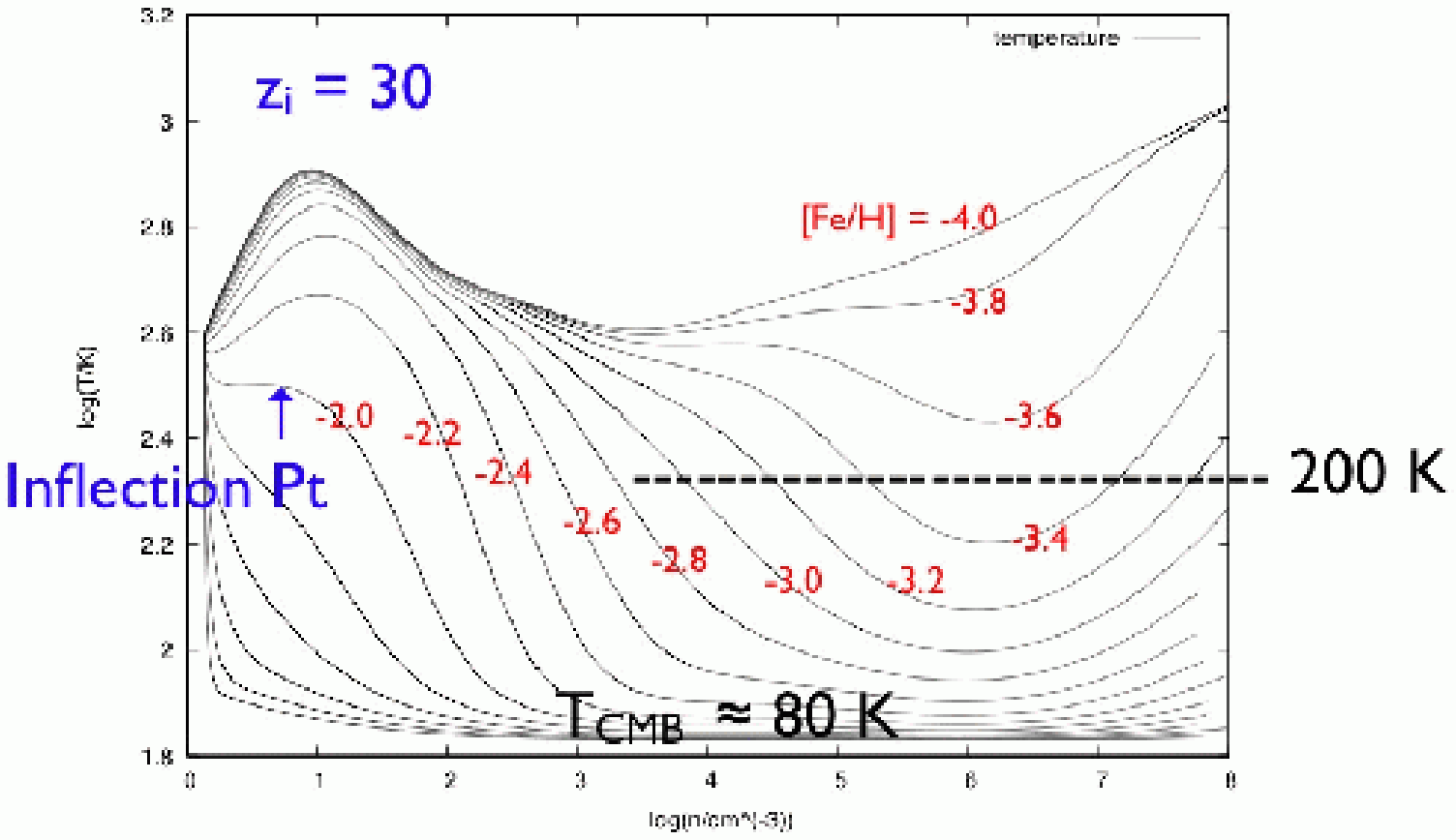}
\caption{Temperature--density ($\log T, \log n)$ evolution of a collapsing 
   gas cloud (Santoro \& Shull 2008) starting with virial conditions of 
   mass, density, and temperature at $z = 30$.  Curves are labelled
   by metallicity $\log (Z/Z_{\odot})$ or [Fe/H].
   The initial rise in temperature comes from adiabatic heating,
   while the subsequent decreases in $T$ are driven by cooling from
   H$^{\circ}$-excited H$_2$ rotational lines and fine-structure
   lines of [O~I], [Si~II], [Fe~II], [C~II].
   The broad temperature minima occur at $n \approx 10^{5.5}$
   to $10^{6.5}$ cm$^{-3}$, near the critical densities for 
   H$^{\circ}$ de-excitation of [O~I], [Si~II], and [Fe~II] 
   fine-structure levels.
   When level populations reach LTE, the volume cooling rate
   ${\cal L} \propto n$ rather than $n^2$, and $T$ begins to rise
   at $n > n_{\rm cr}$.  Radiation and collisions couple the gas and
   fine-structure levels to the CMB ($T_{\rm CMB} \approx$ 70--80 at
   $z = $25--30).  The stellar IMF may be imprinted when
   $T \leq 200$~K and $n \approx 10^6$ cm$^{-3}$. }
\end{figure}


\begin{figure}
\includegraphics[height=8.5cm]{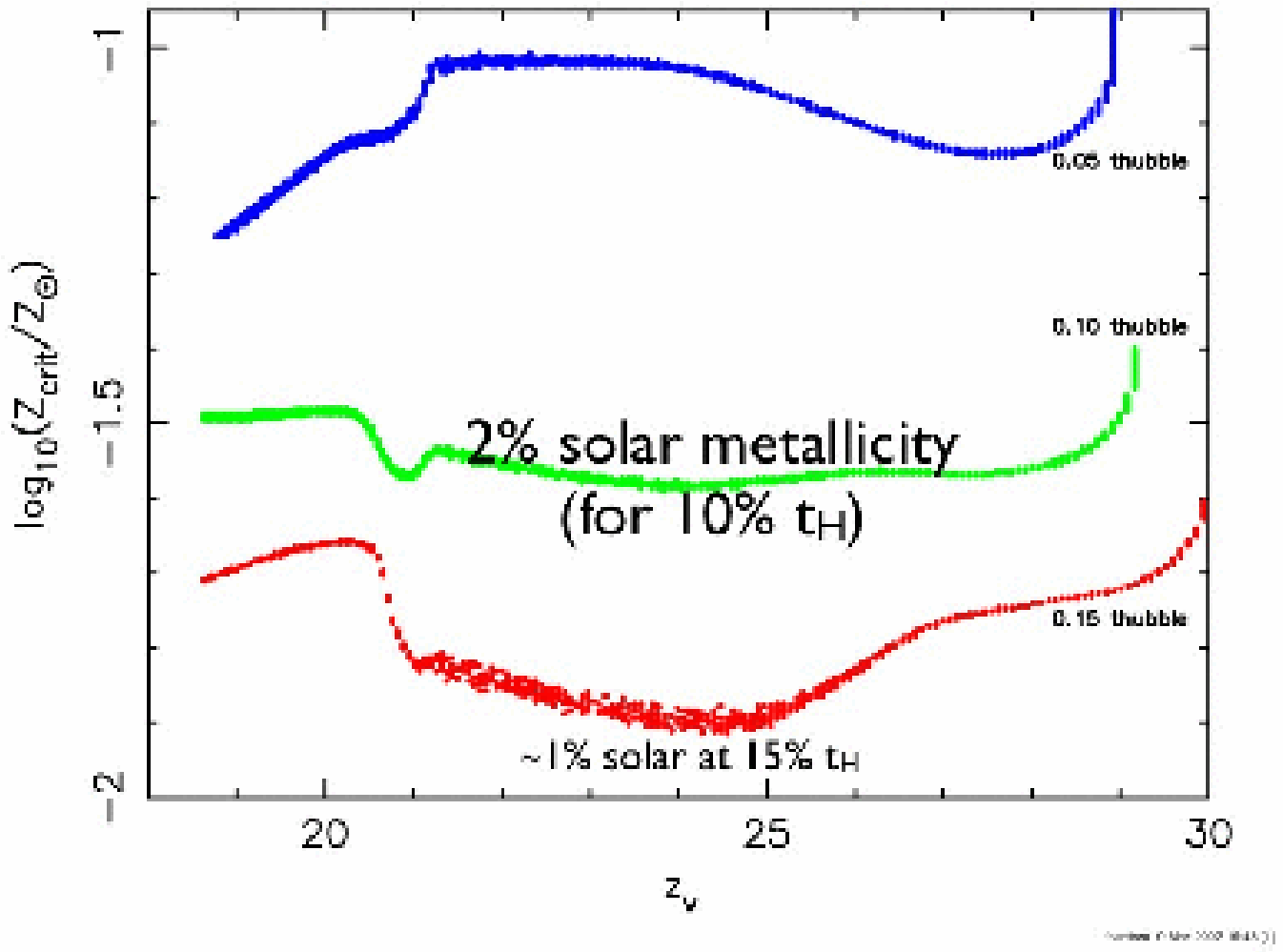}
\caption{Critical metallicity (Santoro \& Shull 2008) 
for collapsing clouds, allowed to cool
for various fractions (5\%, 10\%, 15\%) of the local Hubble time
at $z = $ 20--30.  As shown by the lower two curves, the gas within
virialized halos may need
to reach metallicities as large as $\sim$1--2\% $Z_{\odot}$ in order
to cool in 10--15\% of the local Hubble time, $t_H$. }
\end{figure}

The results of our time-dependent models are shown in Figures 2 and 3, 
with the astrophysical importance summarized in the captions.
Figure 2 shows the thermodynamic $(T,n)$ collapse history at $z=30$.
After an initial rise due to adiabatic heating, the 
temperature turns downward (inflection point at [Fe/H] $\leq -2.0$). 
Once the density exceeds $n \geq 10^6$ cm$^{-3}$ at
$Z > 10^{-3.5} Z_{\odot}$, $T$ is driven toward $T_{\rm CMB}$.
Figure 3 illustrates the dependence of $Z_{\rm crit}$ on the
time allowed to cool, expressed as a fraction of the local Hubble 
time. Most of the cooling time is spent at low densities.
In order to cool in 10--15\% of $t_H$, the gas must reach 
$Z_{\rm crit} \approx 0.01 Z_{\odot}$, until the density reaches
the point ($\sim 10^6$ cm$^{-3}$) of most efficient cooling. 

These far-infrared fine-structure lines dominate the H$_2$
cooling once $Z > Z_{\rm crit}$, and their detection represents
a challenge for FIR and sub-mm astronomy. As shown by
Santoro \& Shull (2006), the strongest lines are [O~I] 63 $\mu$m,
[Si~II] 34.8 $\mu$m, and [Fe~II] 25.99 $\mu$m, which at $z=4$
redshift to 316 $\mu$m, 174 $\mu$m, and 130 $\mu$m.  For 
$(10^8~M_{\odot}) M_8$ of high-density (LTE) gas at 200~K and 
$Z = (0.01 Z_{\odot})Z_{0.01}$, the line luminosities are 
predicted to be 
$L_{\rm line} \approx ([0.8-2.0] \times 10^{41}~{\rm erg~s}^{-1}) 
M_8 Z_{0.01}$.  The luminosity distance at $z \approx 4$ is 
$d_L \approx 10^{29}$ cm, and the expected line fluxes are 
$\sim10^{-21}$ W~m$^{-2}$, within reach of facilities
such as ALMA, SPICA, or SAFIR.  At higher redshifts, these lines
shift into the 350~$\mu$m (sub-mm) window. The fluxes
will probably be lower, owing to the larger $d_L$, lower $Z$, 
and smaller masses of cooling gas.

\section{Observations \& Discussion}

The ``fossil record" of the first stars may be observed in gas-phase
(IGM), as well as in metal-poor halo stars.  
Metallicities $Z \sim 10^{-3}$ Z$_\odot$ are measured from
absorption lines in the Ly$\alpha$ forest at $z \sim 2-5$
(Songaila 2001; Schaye et~al.\ 2003; Pettini et~al.\ 2003;
Simcoe et~al.\ 2004). This mean metallicity, typically inferred from
abundances of C~IV and Si~IV, shows little evolution from $z=5$ to $z=2$.
However, a possible redshift-dependent ionization correction may
conspire to mask any real metallicity evolution.

The origin of the metals observed in the low-density Ly$\alpha$ forest at 
redshifts $z \sim 2-5$ is still under some debate. One view invokes a
nearly uniform pre-enrichment of the intergalactic medium (IGM) produced 
by the first stars at high-redshift (Madau et~al.\ 2001).  The other view 
attributes the origin of the observed metal lines to hot, metal-enriched 
superbubbles located around Lyman-break galaxies (Adelberger et~al.\ 2003).  
Given the difficulties associated with both scenarios, it is important to 
know the amount and volume filling factor of metal-enriched IGM produced 
by primordial stars and galaxies.

As modeled by many groups (e.g., Ricotti, Gnedin, \& Shull 2002, 2007), 
the first sources of metals may come from ``dwarf primordial" (dPri) 
galaxies, with virial temperatures $T_{\rm vir} \leq 20,000$ K and
circular velocities $v_c \leq 20$ km~s$^{-1}$.  
In contrast to more massive galaxies, the 
dark matter (DM) halos of dPri galaxies are too shallow to contain 
much photoionized gas, with temperatures 10,000--20,000 K. During their 
formation, the gas is only heated to temperatures below $10^4$~K, where 
it is unable to cool by atomic hydrogen (Ly$\alpha$) line emission.  
The mass of these halos is $M_{\rm dm} \leq 2 \times 10^{8}$ M$_\odot$ 
at their typical redshifts of formation ($z \geq 10$). These
galaxies rely primarily on the formation of H$_2$ to cool and form 
stars, because metal cooling is negligible as long as
the gas has almost primordial composition.  This situation changes,
after the metallicity rises above $Z_{\rm crit}$,
which could be as large as $0.01 Z_{\odot}$ at halo gas densities 
$n \approx$ 10--100 cm$^{-3}$. As the first stars form, these 
requirements no longer hold, since some gas
is heated above 10,000 K and is polluted with heavy elements.

At low redshifts ($z < 0.4$), ultraviolet spectrographs aboard
{\it Hubble} and {\it FUSE} have made similar measurements of the
gas-phase baryon content and metallicity of the IGM, 
using Ly$\alpha$, Ly$\beta$, and trace metal lines of 
O~VI, C~III, C~IV, etc. (Danforth \& Shull 2005, 2007).  
The current observational sensitivity to metallicity is
$\sim10^{-3.5} Z_{\odot}$ at high redshift (Songaila 2001;
Simcoe et~al.\ 2004) and $10^{-2} Z_{\odot}$ at low redshift
(Danforth \& Shull 2007; Stocke et al. 2007).
Figure 4 shows how these lines can be used to derive
the statistical metallicity ($\sim0.1 Z_{\odot}$) from ratios of
O~VI to H~I.  However, the IGM appears to have widely varying
metal abundances, with differences between ``filaments" and ``voids"
in the ``Cosmic Web" of absorbers.   
If underdense regions of IGM contain some metals, then star formation
in high redshift galaxies may have pre-enriched it (Madau et~al.\ 2001).
If, instead, the metals detected along a line of sight are associated with
nearby bright galaxies at $z \sim 2-5$, the metal distribution is
probably quite inhomogeneous. In this second case, observations are
not probing a minimum floor of metal enrichment produced by the first
galaxies.

The degree of inhomogeneity in the IGM metal distribution IGM is 
not well characterized, although this should change with the installation 
of the powerful {\it Cosmic Origins Spectrograph} (COS) on the 
{\it Hubble Space Telescope} in August 2008.  
The spectroscopic throughput of COS should be over $20\times$ that of
STIS, and the resulting ultraviolet spectra will be used for a wide 
range of observations of the IGM and surrounding galaxies:  
(1) surveys of the baryon content and metallicity of the low-$z$ IGM;
(2) searches for nucleosynthetic patterns among heavy elements;
(3) absorber-galaxy correlation studies, seeking connections between
the IGM and the galaxies responsible for metal enrichment;
(4) Surveys of baryons and metals residing in absorbers within voids.
Preliminary studies of several of these issues with HST and FUSE 
spectrographs is now underway (Danforth \& Shull 2005, 2007; 
Stocke et al.\ 2006; Stocke et al.\ 2007).


\begin{figure}
\includegraphics[height=15cm]{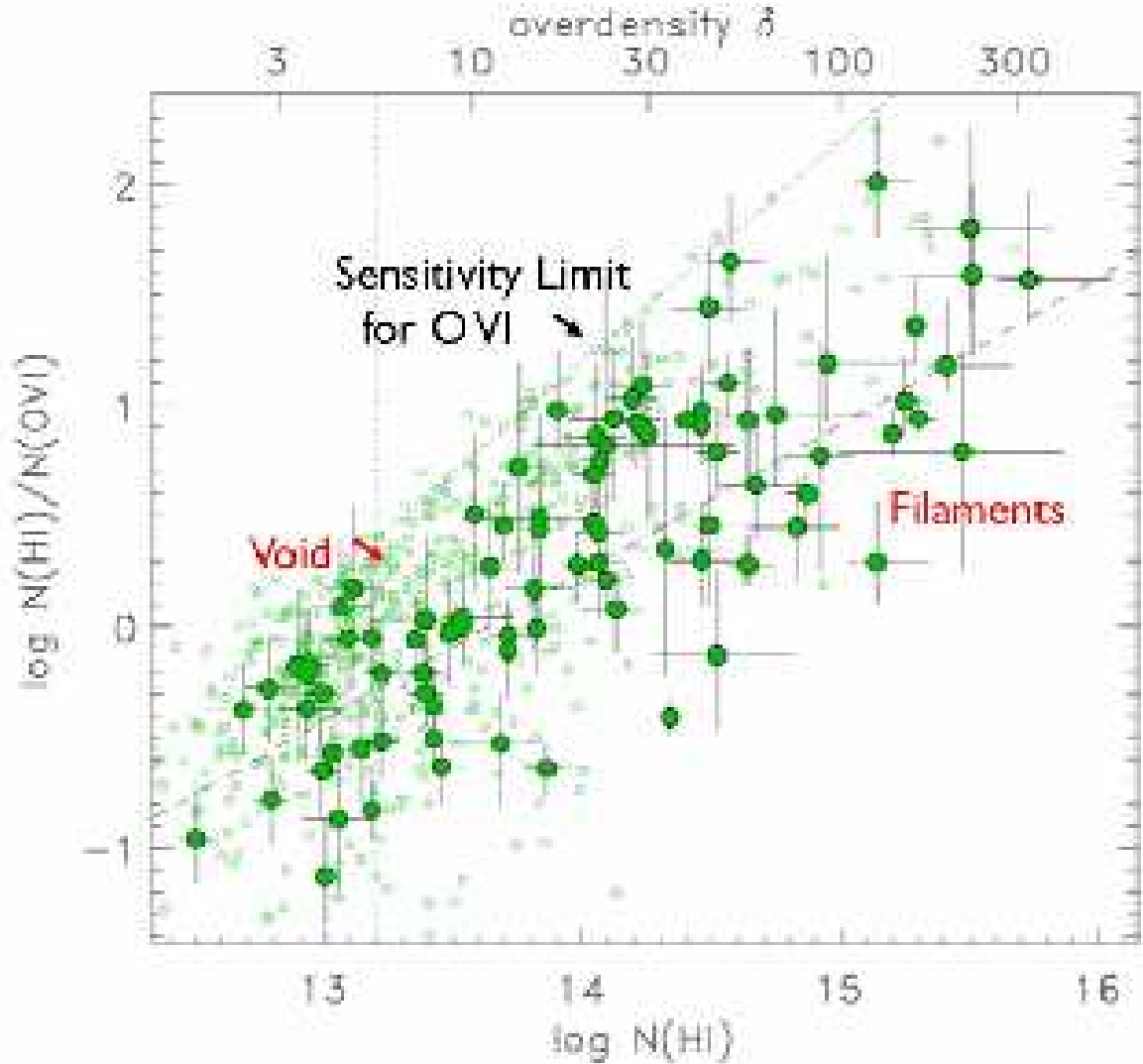}
\caption{The ``fossil record" of the first stars may be studied 
in the gas phase at $z < 0.4$, through quasar absorption lines. 
This plot shows the ``multiphase ratio" of O~VI and H~I absorbers, 
measured by {\it Hubble Space Telescope} and the {\it Far Ultraviolet
Spectroscopic Explorer} (Danforth \& Shull 2005, 2007).  The ratio,
N(H~I)/N(O~VI) varies from high column density filaments in the 
Cosmic Web to low-column gas in voids, with 
N$_{\rm HI} < 10^{14}$ cm$^{-2}$.  Our
surveys and abundance analyses show that low-$z$ filaments have
mean metallicities $\sim0.1 Z_{\odot}$, whereas no metals have
been detected in voids, down to limits of $0.02Z_{\odot}$ (Stocke
et al.\ 2007).  The major conclusions of this study are:
(1) the filaments have increased their metallicity by factors 
of 30--100 from $z \geq 3$ to the present; and (2) the IGM in voids
may contain low-metallicity pockets at $Z < 0.01 Z_{\odot}$. } 
\end{figure}




\begin{theacknowledgments}
 This work has been supported by the Astrophysical Theory program
at the University of Colorado, through NASA grant NNX07-AG77G, and
through grants from the Space Telescope Science Institute and 
from NASA for far-UV studies with the FUSE satellite. 
\end{theacknowledgments}



\bibliographystyle{aipproc}   

\bibliography{sample}

\IfFileExists{\jobname.bbl}{}
 {\typeout{}
  \typeout{******************************************}
  \typeout{** Please run "bibtex \jobname" to optain}
  \typeout{** the bibliography and then re-run LaTeX}
  \typeout{** twice to fix the references!}
  \typeout{******************************************}
  \typeout{}
 }

\end{document}